\documentclass{ems-mag-author}
\usepackage{orcidlink}
\begin{document}

\doi{???}
\issue{???}

\title{Fraudulent Publishing in Mathematics: A European Call to Action and How Information Infrastructure Can Help}
%Alphabetic order
\author{\orcidlinki{Moritz Schubotz}{0000-0001-7141-4997}, \orcidlinki{Jan Philip Solovej}{0000-0002-0244-1497}}

\maketitle

\begin{abstract}
The IMU--ICIAM working group's new report on Fraudulent Publishing in the Mathematical Sciences documents how gaming of bibliometrics, predatory outlets and paper-mill activity are eroding trust in research, mathematics included. This short EMS note brings that analysis home to Europe.
We urge readers to recognise the warning signs of fraudulent publishing, to report serious irregularities so that they can be investigated and sanctioned, and to reflect critically on their own editorial and reviewing practices. We then sketch why Europe is well placed to lead a structural response: a decade of policy development on open science; mature infrastructures for data, software and scholarly communication; and new capacity for community-led diamond open access. Finally, we outline developments towards non-print contributions across member countries including the growth of formal proofs (e.g.\ with \emph{Lean} and \emph{Isabelle}) and we highlight the role of \emph{zbMATH Open} as a European quality signal that can help editors, reviewers and authors steer clear of problematic venues.
\end{abstract}

\section{A Call to Action}
The core message of the IMU--ICIAM report~\cite{imu-iciam, imu-iciam2} is sobering: when numbers become targets, they invite manipulation. In fields like mathematics, with fewer papers and lower citation volumes, a handful of strategic behaviors (excessive self-citation, cartelized referencing, ``special-issue mills'') can strongly skew metrics, rankings, and careers.
Europe is not immune.
The business shift to APC-funded mega-journals and the normalisation of quantitative research assessment have created perverse incentives. We now see:
\begin{description}
    \item [Volume-over-value publishing:] large fleets of themed special issues; light or inconsistent peer review; guest-editor networks oriented around throughput.
    \item[Citation gaming:] editorial pressure to add citations post-accep\-tance; reciprocal referencing; opaque ``recommended citations''.
    \item[Paper-mill leakage]: templated articles with ``tortured phrases'', fabricated references, or recycled figures that sometimes pierce editorial checks.
    \item[Downstream effects:] hiring and promotion tilted toward countable surrogates; university rank chasing; loss of trust among collaborators and early-career colleagues.
\end{description}

This is fixable, but only if we act where Europe has natural leverage: community-owned infrastructures, curated indexing, open-science standards, and editorial leadership.
Fraudulent publishing damages the literature, wastes researchers' time, and distorts incentives. While most mathematicians act in good faith, experience shows that small distortions of bibliometric targets can escalate into serious manipulation.
The first responsibility is therefore individual~\cite{imu-iciam2}: read the papers you evaluate; check venues before submitting or accepting an editorial role; and, as a reviewer or editor, insist on transparent authorship and sound referencing practices. When you encounter serious irregularities (plagiarism, paper-mill products, manufactured citations, or identity misuse), report them through journal procedures; if necessary, raise concerns with institutions or recognised community channels. 

Editors and editorial-board members have special duties. If commercial pressure, guest-editor overload, or systematic citation manipulation erodes standards, resigning (and explaining why) is sometimes the only way to protect the record. This is uncomfortable, but it is often decisive. Ultimately, the community will only repair incentives if we act early and visibly. The aim is not policing for its own sake; it is to preserve trust in our literature and to protect younger colleagues from short-term temptations that can harm careers and institutions in the long run.

\smallskip
Within the Society, the EMS Committee for Publications and Electronic Dissemination (PED) has issued guidance on sound publishing practice and predatory outlets, which complements the IMU Committee on Publishing (CoP) recommendations~\cite{ems-ped,ems-predatory}.

\section{Why Europe can Lead}
Europe has aligned policy and infrastructure that make com\-munity-go\-verned publishing feasible at scale. In May~2023 the Council of the European Union called for immediate and unrestricted open access to publicly funded research, for transparency in publishing contracts and costs, and for robust quality and integrity safeguards~\cite{councilpress,council8827}. These conclusions recognise the importance of non-profit, scholar-led models (often called \emph{diamond} open access) as part of a sustainable ecosystem.

\smallskip
The European Open Science Cloud (EOSC) provides a federated fabric for data, software and services; importantly for our community, the European Mathematical Society (EMS) participates in the EOSC Association as an \emph{observer}, ensuring that mathematical perspectives inform EOSC's evolution~\cite{eoscems}. For journals and platforms, Europe's \emph{European Diamond Capacity Hub} (EDCH) has emerged as a coordination point: it gathers funders, infrastructures and communities to produce shared guidance, a common access point to services, and a living map of the diamond landscape~\cite{edch}. The Horizon Europe project DIAMAS underpins this work with practical recommendations for institutional publishing in Europe and with an inventory of services and requirements~\cite{eua-diamas,cordis-diamas}.

\smallskip
These pan-European efforts dovetail with national capacity. In Germany, SeDOA (the {Servicestelle Diamond Open Access}) acts as a national node that curates information, offers guidance to journals, and connects institutions and funders with the European conversation~\cite{sedoa}. In the Netherlands, \emph{Openjournals.nl}, provides an integrated diamond platform used by many scholarly journals across disciplines~\cite{openjournals}. Such initiatives make it realistic for editorial boards to migrate away from risky models without sacrificing visibility or service quality.

EMS itself has taken a leading role in shaping open access publishing. In 2019, the Society transformed its publishing house into \emph{EMS Press}, today operating under the Subscribe-to-Open (S2O) model~\cite{ems-s2o} for its journals. While EMS Press S2O publications are only made freely available to readers if sufficient subscriptions are secured, the model is free for authors, applies an open access–compliant licence, and lets the EMS or other journal-owing societies of institutions claim the publication rights. EMS Press is thus organised as a professional publisher, but with EMS as its sole shareholder its orientation is not towards maximising profit but towards serving the scientific community.

This approach resonates with the definition of \emph{Diamond Open Access} articulated by the German Servicestelle Diamond Open Access (SeDOA): ``In Diamond Open Access journals, quality-assured articles are published free of charge for authors and readers under an open access–compliant licence. The journal or title rights belong to the scholarly community. Diamond Open Access journals serve academic communication and non-commercial purposes~\cite{sedoa}.'' In this sense, EMS Press is a distinctive hybrid: while operating commercially in form, its governance and S2O model place it firmly in the landscape of community-oriented publishing, ensuring that quality and openness remain at the centre.

\section{Member-country Developments: Data, Software and Formal Proofs}
Open science in mathematics is not only about access to articles. Reproducibility depends on sharing \emph{data}, \emph{software} and, increasingly, \emph{formal proofs}. Several member countries are investing in field-specific infrastructures. In Germany, the Mathematical Research Data Initiative (MaRDI) within the National Research Data Infrastructure (NFDI) develops standards for mathematical research data, verified workflows, and services, including a portal that interlinks objects, algorithms, software and literature while implementing the FAIR principles~\cite{mardi-nfdi}.

On the software side, \emph{swMATH} (now integrated into zbMATH Open) identifies and interlinks mathematical software with the literature, and is being connected to persistent-identifier graphs used in EOSC to improve discovery and credit~\cite{swmath-zb,swmath-ems}. This makes it easier for editors and referees to check that computational claims are supported by citable software artefacts.

Formal proofs are also gaining traction. The COST Action \emph{EuroProofNet} brought together more than 40 countries to improve the interoperability and usability of proof assistants and libraries across Europe~\cite{europroofnet}. Meanwhile, communities around \emph{Lean}/\emph{mathlib} and \emph{Isabelle} continue to expand, with training schools and research programmes across the continent~\cite{lean-mathlib,isabelle}. For editors and referees, this might matters in practice: when feasible, formally verified arguments may reduce the attack surface for paper-mill text, fabricated claims, and accidental error.

\section{European Quality Signals: the Role of \emph{zbMATH Open}}
Located in Europe and co-edited by the EMS, \emph{zbMATH Open} curates the mathematical literature with post-publication reviews and rich interlinking to data and software. Crucially for quality assurance, indexing may be \emph{cover-to-cover or in parts}, and indexing decisions are reviewed over time~\cite{zb-faq}. In consequence, journals whose practices deteriorate can be excluded from further coverage, and in multidisciplinary venues only mathematically relevant items are indexed. For editors, authors and evaluators, this provides a pragmatic signal: if a venue is no longer covered or is only partially indexed, caution is warranted. In parallel, recent developments at \emph{zbMATH Open}, for example the display of OA and licensing information at article level and tighter integration of software metadata, help the community to verify provenance, reuse conditions and computational support~\cite{zb-oa,swmath-zb}.

\section{What EMS members can do now}
Mathematicians are central stewards of the literature. In Europe we benefit from aligned policy and strong community infrastructures, but their effect depends on our everyday choices. First, when you act as an editor or reviewer, hold the line on quality and transparency, and use community services like \emph{zbMATH Open} and \emph{swMATH} to check provenance and software. Second, when a venue's behaviour raises concerns, \emph{say so}: report, withdraw support, and encourage colleagues to move to safer homes; European diamond platforms now make this transition viable. Third, invest in openness by sharing data, code and, when appropriate, formal proofs; this narrows the space for low-quality and fraudulent work to hide. Europe has the policy framework and the infrastructure; if we align our publishing choices with these tools, we can rebuild incentives that reward \emph{mathematics}, not metrics.

\emph{Acknowledgements:} This paper summarises the companion report \emph{Fraudulent Publishing in the Mathematical Sciences} \cite{imu-iciam, imu-iciam2} and aligns its recommendations with current European open-science infrastructures. The authors thank colleagues in EMS in particular the EMS Committee for Publications and Electronic Dissemination (PED), IMU in particlar the Committee on Publishing (CoP) and ICIAM for comments and discussions. The text of this articles was improved using advanced grammar correction and large language models.
\bibliographystyle{plain}

\end{document}